\documentclass[fleqn,usenatbib]{mnras}

\usepackage{newtxtext,newtxmath}

\usepackage[T1]{fontenc}
\usepackage{ae,aecompl}

\usepackage{graphicx}	
\usepackage{amsmath}	
\usepackage{wrapfig}
\usepackage{float}
\usepackage{lipsum}
\usepackage{subcaption}
\usepackage{mwe}

\title[Ionized Gas in IC 1459]{Deciphering the Origin of Ionized Gas in IC 1459 with VLT/MUSE}

\author[C. R. Mulcahey et al.]{
C. R. Mulcahey,$^{1,4}$\thanks{E-mail: mulca23c@mtholyoke.edu}
L. J. Prichard,$^{2}$\thanks{E-mail: lprichard@stsci.edu}
D. Krajnovi{\'c}$^{3}$
and R. A. Jorgenson$^{4}$
\\
$^{1}$Mount Holyoke College, 50 College St, South Hadley, MA 01075, USA\\
$^{2}$Space Telescope Science Institute, 3700 San Martin Drive, Baltimore, MD 21218, USA\\
$^{3}$Leibniz-Institute for Astrophysics, An der Sternwarte 16, 14482 Potsdam, Germany\\
$^{4}$Maria Mitchell Observatory, 4 Vestal Street, Nantucket, MA 02554, USA
}

\date{Accepted XXX. Received YYY; in original form ZZZ}

\pubyear{2020}

\begin{document}
\label{firstpage}
\pagerange{\pageref{firstpage}--\pageref{lastpage}}
\maketitle

\begin{abstract}
IC 1459 is an early-type galaxy (ETG) with a rapidly counter-rotating stellar core, and is the central galaxy in a gas-rich group of spirals. In this work, we investigate the abundant ionized gas in IC 1459 and present new stellar orbital models to connect its complex array of observed properties and build a more complete picture of its evolution. Using the Multi-Unit Spectroscopic Explorer (MUSE), the optical integral field unit (IFU) on the Very Large Telescope (VLT), we examine the gas and stellar properties of IC 1459 to decipher the origin and powering mechanism of the galaxy’s ionized gas. We detect ionized gas in a non-disk-like structure rotating in the opposite sense to the central stars. Using emission-line flux ratios and velocity dispersion from full-spectral fitting, we find two kinematically distinct regions of shocked emission-line gas in IC 1459, which we distinguished using narrow ($\sigma$ $\leq$ 155 km s$^{-1}$) and broad ($\sigma$ $>$ 155 km s$^{-1}$) profiles. Our results imply that the emission-line gas in IC 1459 has a different origin than that of its counter-rotating stellar component. We propose that the ionized gas is from late-stage accretion of gas from the group environment, which occurred long after the formation of the central stellar component. We find that shock heating and AGN activity are both ionizing mechanisms in IC 1459 but that the dominant excitation mechanism is by post-asymptotic giant branch stars from its old stellar population.
\end{abstract}

\begin{keywords}
galaxies: individual: IC 1459, -- galaxies: elliptical, --
galaxies: evolution, -- galaxies: kinematics and dynamics, -- galaxies: nuclei, -- galaxies: structure
\end{keywords}

\section{Introduction} 
\label{sec:intro}

Early-type galaxies (ETGs) are typically characterised as quiescent, having an old stellar population, little gas and dust, and being devoid of spiral arms and structure. A feature unique to ETGs are kinematically distinct cores (KDCs), which occur when there is a rapid change of the position angle of the kinematic axis of more than 30$^{\circ}$ and the rotational velocity approaches zero in the transition region \citep{Krajnovic2011}. The first detection of KDCs in ETGs was made by \cite{Efstathiou1982} using long-slit spectroscopy to study stellar kinematics. Subsequent studies used a similar approach and detected additional KDCs \citep{Bender1988b, Carollo1994, Franx1988, Jedrzejewski1988}. Long-slit spectroscopy can be costly for large galaxy samples and only produces one spectrum per galaxy from which physical properties of the whole system are inferred. The advent of wide-field, high-resolution integral field spectroscopy (IFS) has revolutionised our understanding of ETGs by enabling more detailed investigations (see \citealt{Cappellari2016} for a review). Instead of obtaining one spectrum of light per galaxy, integral field units (IFUs) allow a galaxy to be viewed in three dimensions (3D), where every spatial pixel (spaxel) across the surface of a galaxy contains a full spectrum of light.

The SAURON \citep{Bacon2001} survey \citep{deZeeuw2002} was the first IFS study to spatially resolve stellar kinematics, ionized gas, and stellar populations, and did so for a sample of 48 nearby ETGs. Their spatially resolved kinematics \citep{Emsellem2004} qualitatively separated ETGs into two classes: irregular, slow rotating, high-mass (M$_{\rm crit}$ $\gtrsim$ 2 $\times$ 10$^{10}$ M$_{\rm \odot}$) ETGs ($\sim$ 15$\%$ of population) and regular, fast rotating, low-mass ETGs \citep{Emsellem2007}, which resemble disk models \citep{Krajnovic2008}. The ATLAS$^{\rm 3D}$ survey \citep{Cappellari2011a} combined IFS data with theoretical modelling, which led to additional sub-classifications of slow rotators \citep{Krajnovic2011}. Although KDCs are rare in ETGs ($\sim$ 7$\%$), they are relatively common in slow rotators \citep[$\sim$ 42$\%$;][]{Krajnovic2011}.

The stellar and gas kinematics from each spaxel of IFS data provides a wealth of information that can be used to understand a galaxy's inherent properties and evolution in greater depth. Investigating stellar and gas components in isolation reveals unique evolutionary processes and is made possible by developments in full-spectral fitting codes and supporting spectral libraries and stellar population synthesis models \citep[e.g.,][]{Cappellari2004, Cappellari2017, Vazdekis2010, Vazdekis2016}. Studying the stellar component, for example, discerns whether a galaxy has been constructed from gas-poor or gas-rich merging events \citep[e.g.,][]{Cappellari2013b, Emsellem2011, Naab2014}. However, the stellar component does not reveal the complete evolutionary history of a galaxy because a notable percentage of ETGs exhibit ionized gas in the emission-lines of optical spectra \citep[40-80$\%$; e.g.,][]{Caldwell1984, Phillips1986, Kim1989, Bunson1993, Goudfrooij1994, Macchetto1996, Sarzi2006, Sarzi2009}. 

Emission-line gasses reveal how gas accretion or gas-rich mergers have influenced the assembly of the galaxy \citep[e.g.,][]{BarreraBallesteros2015, Davis2011, Sarzi2006} and can be used to understand the mechanisms behind star formation \citep[SF; e.g.,][]{Alatalo2011, Cheung2016, Ho2016}. Emission-line ratios, which are used extensively in the Baldwin-Phillips-Terevich (BPT) diagrams \citep{Baldwin1981, Veilleux1987} and the WHAN (equivalent width of H$\alpha$ versus [NII]/H$\alpha$) diagram \citep{CifFernandes2011}, are key diagnostics used to estimate whether the principal power source of galaxies is from SF or from the central active galactic nucleus (AGN). With IFU data, the emission-line ratios of these diagrams can be spatially resolved, revealing multiple ionization mechanisms across the surface of a galaxy. Moreover, IFU data can unveil extended shocks from outflows, galactic winds, and tidally induced shocks \citep[e.g.,][]{MonrealIbero2006, MonrealIbero2010, Farage2010, Rich2010, Rich2011}.

To hone our understanding of the formation and evolution of galaxies with KDCs, we investigate IC 1459, a prototypical, massive \citep[M$\sim$ 4-6 $\times$ 10$^{11}$ $M_{\rm \odot}$;][]{Cappellari2002,Samurovic2005} ETG with a counter-rotating core. IC 1459 is a local \citep[D = 30.3 $\pm$ 4.0 Mpc;][]{Ferrarese2000}, bright \citep[M$_{\rm v}$ $\simeq$ -22.3;][]{Paturel1997} E3 ETG and it is the central galaxy of a gas-rich group of 11 galaxies, which mainly consists of spirals \citep[group number 15;][]{Huchra1982, Serra2015}. The galaxy shows clear evidence for past group interactions like tidal features in deep images \citep{Malin1985} and shells at large radii \citep{Forbes1995}. The outer regions of the galaxy and central ionized gas component rotate in one direction at V$_{\rm max}$ $\sim$ 45 $\pm$ 8 km s$^{-1}$ while the central stellar component rapidly counter rotates at V$_{\rm max}$ $\sim$ 170 $\pm$ 20 km s$^{-1}$ \citep{Franx1988}. 

IC 1459 is a gas-rich ETG containing molecular gas \citep{Prandoni2012}, and ionized gas in its core \citep[e.g.,][]{Franx1988, Phillips1986}. Neutral HI gas, which was most likely stripped away from NGC 7418 through tidal interactions \citep{Oosterloo2018, Iodice2020}, is abundant in the intra-group environment \citep{Serra2015, Saponara2018}. IC 1459, itself, does not contain HI gas \citep{Serra2015}, though the galaxy's ionized gas can be linked to the surrounding HI clouds \citep{Saponara2018}. IC 1459 hosts a bright radio source \citep[PKS 2254–367; ][]{Tingay2015} and \cite{Cappellari2002} measured the mass of the central black hole to be between $\sim$ 3.5 $\times$ 10$^8$ M$_{\odot}$ (from gas kinematics) and 2.6 $\pm$ 1.1 $\times$ 10$^9$ M$_{\odot}$ (from stellar kinematics). The galaxy is also characterised as a Low Ionisation Nuclear Emission-line Region \citep[LINER; ][]{Phillips1986, VerdoesKleijn2000, Annibali2010}.

A recent optical (MUSE: Multi-Unit Spectroscopic Explorer, \citealt{Bacon2010}) and infrared (KMOS: K-band Multi-Object Spectrograph, \citealt{Sharples2013}) IFU study of IC 1459 investigated the evolution of the galaxy focusing on the stellar populations, initial mass function (IMF), and stellar kinematics \citep{Prichard2019}. The stellar velocity dispersion revealed that IC 1459 is dynamically hot along the major axis, which is indicative of counter-rotating stellar components extending throughout the galaxy beyond the core. The stars show a radially flat and bottom-heavy IMF and old stellar ages throughout the galaxy, further supporting the theory that it comprises smooth co-spatial counter-rotating populations. However, orbital modelling is needed to fully understand if this is the case for IC 1459.  We know that the counter-rotating stellar component formed at early times and therefore the stars alone do not give the full evolutionary picture of this gas-rich ETG. The origin of IC 1459's abundant gas remains unknown and the excitation mechanisms of its ionized gas is undetermined. We are also yet to understand the relationship, if any, between the gaseous and stellar components of the galaxy as they exist today. 

To help us understand the evolutionary history of IC 1459, we build upon the work of \cite{Prichard2019} and use the existing MUSE data to study the galaxy's ionized gas in detail. We also construct dynamical models to investigate the internal orbital structure, the nature of the KDC, and its potential link with the emission-line gas. This paper is organised as followed: Section \ref{sec:Observations and Data Reduction} outlines the MUSE observations and data reduction, Section \ref{sec:Data Analysis} details the spectroscopic analysis of the MUSE cube. In Section \ref{sec:The ionized Gas of IC 1459} we present the resulting kinematic maps of IC 1459's ionized gas, Section \ref{sec:dyn} details new orbital models for the stellar population in IC 1459, and we  discuss and conclude our findings in Section \ref{sec:Conclusion}.

\begin{figure*}
\centering
\includegraphics[width=\textwidth]{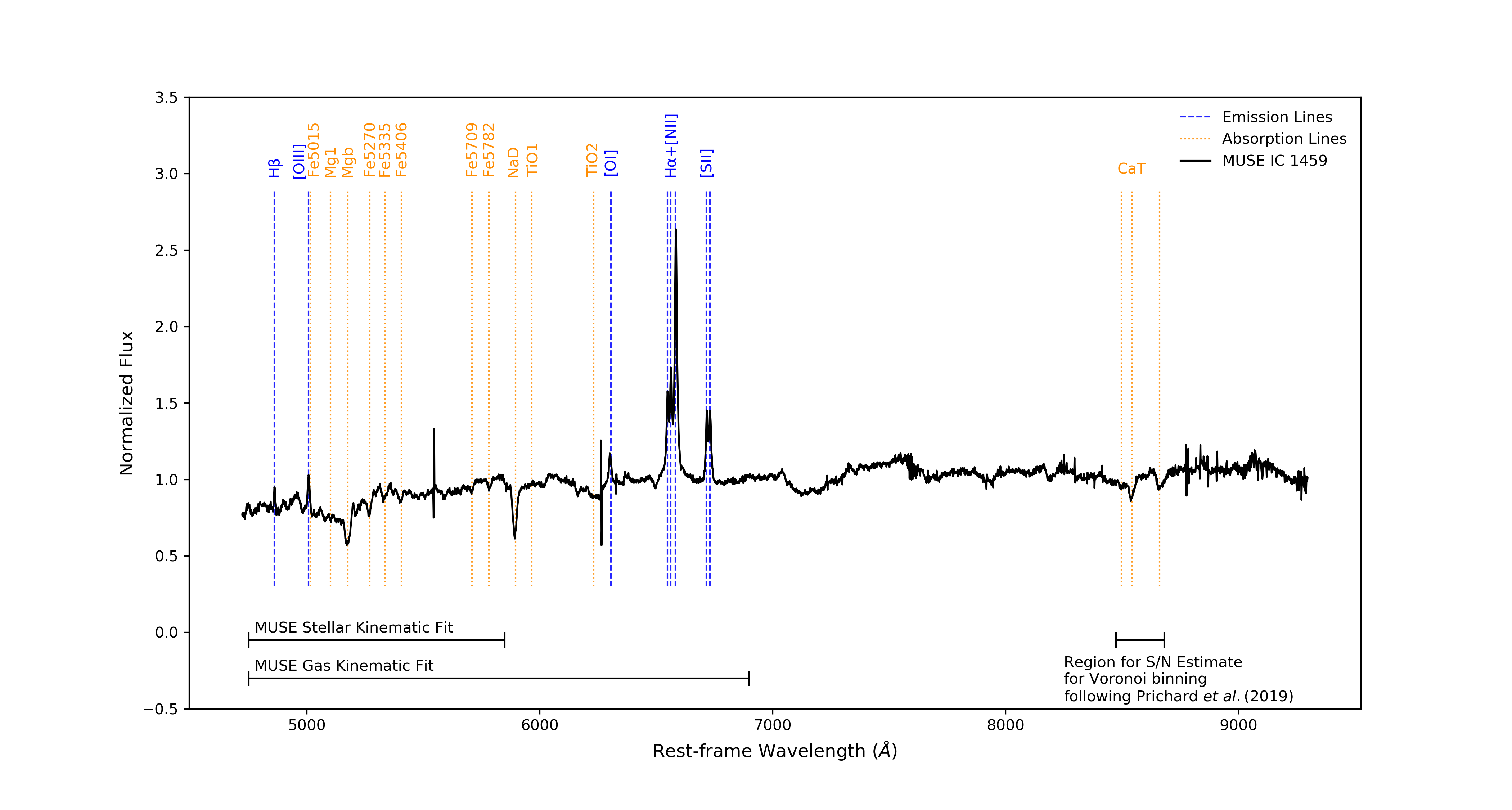}
\caption{Spectrum extracted from the central spaxel of IC 1459. Gas emission lines (blue dashed lines) and stellar absorption lines (orange dotted lines) are shown. All unlabelled spectral features are sky-line contamination. Below the spectrum, we indicate the kinematic fit ranges for extracting gas and stellar kinematics, and the region where conservative S/N values were estimated for Voronoi binning in-line with \citealt{Prichard2019}. See Section \ref{subsec:Voronoi Binning}. \label{fig:medianSpec}}

\end{figure*}

\begin{figure*}

\includegraphics[width=\textwidth, trim=0 0 0 0, clip]{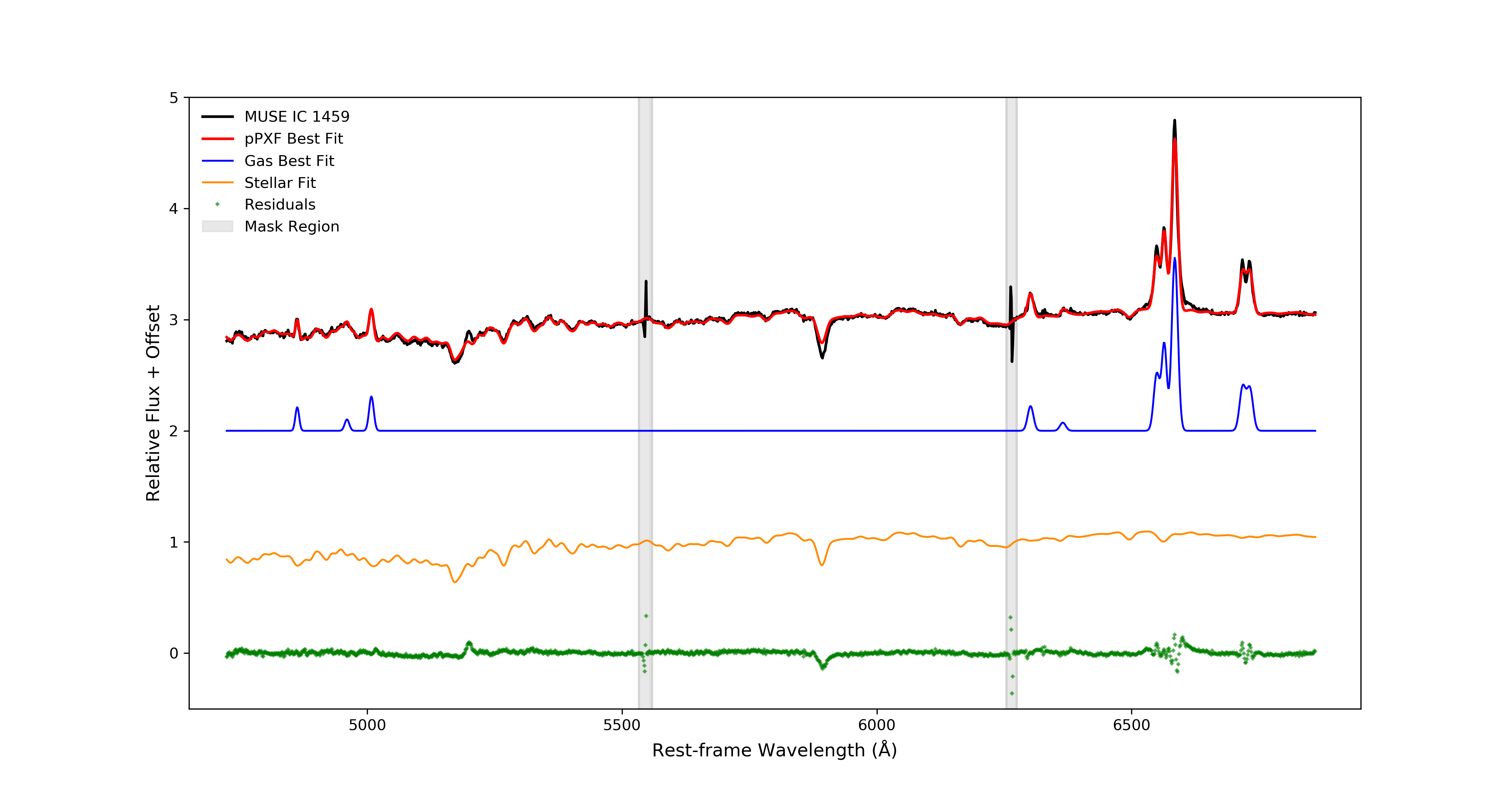}\vspace{1mm}
\caption{pPXF fit to the spectrum of the central spaxel of IC 1459 over the gas kinematics fitting region. The MUSE spectra (black), best fit (red), gas emission component (blue), stellar component (orange), and residuals (green) are shown. The grey shaded regions are locations of bright sky lines that were masked.}
\label{fig:ppxfExample}
\end{figure*}

\section{Observations and Data Reduction}
\label{sec:Observations and Data Reduction}

The MUSE data used in this work is the same cube presented and prepared as described in \cite{Prichard2019}. We briefly summarise the observations and reduction of the MUSE cube here.
MUSE is mounted on the European Southern Observatory's (ESO's) Very Large Telescope (VLT) in Paranal, Chile. It is a large field-of-view (FOV; $1^{\prime} \times 1^{\prime}$) panchromatic optical ($\sim$ 4600--9300 $\text{\AA}$) instrument with a Wide Field Mode (WFM) resolving power ranging R = 1770--3590 and a pixel scale of 0.2$^{\prime\prime}$. The MUSE data of IC 1459 were taken October 14, 2014 (program ID 094.B-0298(A), P.I. C. J. Walcher) with total integration time of 368 s and seeing $\sim$ 1.3--1.6" using the \texttt{MUSE\_wfmnoao\_obs\_genericoffset} observing template. The MUSE observations were taken during bright time (high Moon illumination) so have increased photon noise.

Reduced MUSE data is accessible on the ESO Science Archive\footnote{\url{http://archive.eso.org/cms.html}}. The data were reduced using the \textsc{MUSE-1.6.1} pipeline and were published as a reduced data cube on the {22}$^{\rm nd}$ June 2016. The reduced MUSE data cube of IC 1459 still contained strong contamination from bright sky lines. An additional sky subtraction was executed by subtracting a spectrum from the edge of the MUSE cube, that was dominated by sky, from the rest of the spaxels\footnote{The subtraction of the pseudo sky and cube cropping was done by Joshua Warren and is described in his DPhil thesis.}. The cube was then cropped to remove regions strongly affected by sky and that had low signal-to-noise (S/N). The resulting cube is the central 30$^{\prime\prime} \times$ 30$^{\prime\prime}$, 150 $\times$ 150 pixels centred on IC 1459's core. Residual sky features are still visible across the cube, especially in the red end of the spectra as demonstrated in Figure \ref{fig:medianSpec}. For our analyses of the gas emission (Section \ref{subsec:Muse Kinematics}), we use the bluest part of the spectrum, least affected by residual sky features, and masked bright sky lines. 

\section{Data Analysis}
\label{sec:Data Analysis}

\subsection{Voronoi Binning}
\label{subsec:Voronoi Binning}

Spectra from the MUSE data cube were used to measure the stellar and gas kinematics of IC 1459 (Figure \ref{fig:ppxfExample}). Because the quality and S/N of the spectra varied across the MUSE cube, we used the Voronoi binning method of \cite{Cappellari2003} that adaptively bins data into regions of constant S/N ratio. The \texttt{VORONOI\_2D\_BINNING}\footnote{\url{http://www-astro.physics.ox.ac.uk/~mxc/software/}} software needed a signal and noise for each spaxel over the MUSE cube. To get initial estimates of the S/N of each spectrum for creating the Voronoi bins, we opted to use the calcium triplet (CaT), as used in \cite{Prichard2019}, so that the results of the two studies could be accurately compared. We also required the larger bins and smoother variations for the stellar orbital modelling presented in Section \ref{sec:dyn}. As we fit both the stars and gas within each bin, we wanted to accurately compare both components across the galaxy by extracting their properties within the same bins. The S/N values were determined by fitting the continuum region of the CaT \citep[as defined by ][]{Cenarro2001}. Signal values were estimated from the average continuum level around the CaT and noise values were estimated from the standard deviation of these continuum data. The MUSE cube was binned to S/N = 20, which produced 4329 unique bins. 

Although the CaT feature is in the noisier red end of the spectrum and the binning is therefore more conservative than using the bluer end of the spectrum, the resulting bins have only one or two spaxel size at the centre of the galaxy and these get progressively larger towards the lower S/N outskirts as expected, so we deemed these bins more than sufficient for our analysis of the resolved properties of the gas. To ensure that using the CaT feature rather than a cleaner feature toward the blue end of the spectrum did not affect our results, we also measured the S/N from continuum regions around Mg\textit{b}. We found that the kinematics and results extracted from the Mg\textit{b} bins were the same, and that the only difference was the finer bin sizes for the same S/N threshold as CaT. Increasing the Mg\textit{b} S/N binning threshold to produce roughly the same number of bins as for CaT produced a comparable bin distribution. As we required the coarser bins anyway for the orbital modelling, we stuck with using the CaT feature so we could directly compare results with \cite{Prichard2019}. Normalising and taking the median of the spectra in each Voronoi bin produced spectra to analyse above a threshold S/N value across the entire MUSE cube.

\begin{figure*}
\centering
\includegraphics[width=\textwidth, trim=0 0 0 0, clip]{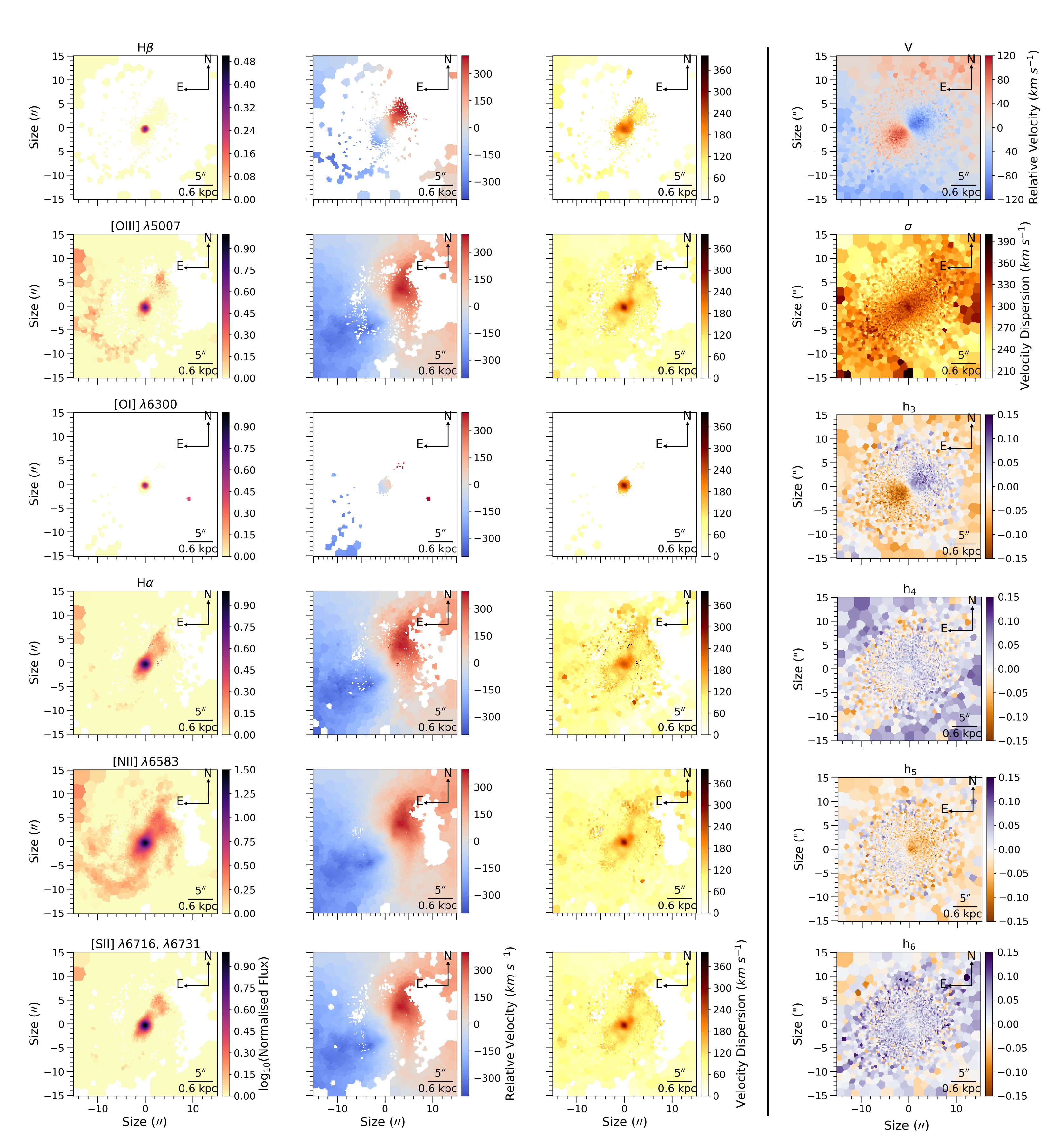}\vspace{1mm}

\caption{\textit{Left of vertical black line:} Maps of normalised emission-line flux (left column), relative velocity (middle column), velocity dispersion (right column) for all bins with A/rN>3 for each feature as measured with pPXF. The emission-line features from top to bottom are: H$\beta$, [OIII]\_d $\lambda$5007,  H$\alpha$, [NII]\_d $\lambda$6583, [OI]\_d $\lambda$6300, and [SII] $\lambda$6731, 6731. \textit{Right of vertical black line:} Maps of the stellar LOSVD moments of IC 1459 as measured with pPXF. \textit{Top to bottom:} stellar relative velocity, velocity dispersion, Gauss-Hermite coefficient $h_3$, $h_4$, $h_5$, and $h_6$.
\label{fig:GasMaps}}
\end{figure*}

\begin{figure*}
    \centering
    \includegraphics[width = \textwidth]{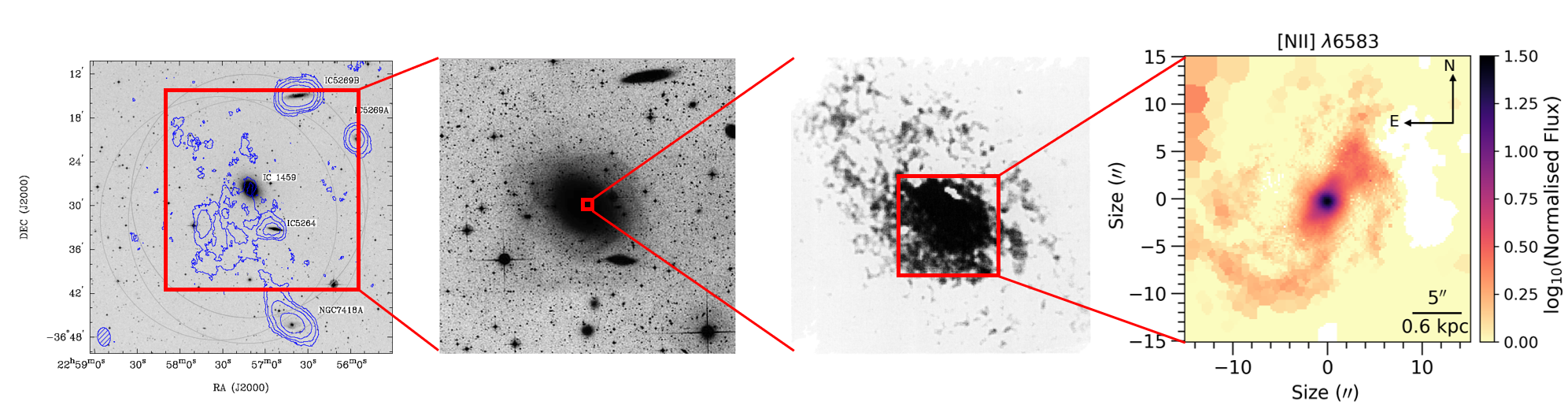}
    \caption{Images of IC 1459 at progressively smaller radii. From left to right: HI map of the IC 1459 group ($\sim$ 40$^{\prime}$ across) from \citealt{Saponara2018}, deep optical image of IC 1459 ($\sim$ 28$^{\prime}$ across) from \citealt{Malin1985}, H$\alpha+$[NII] image of IC 1459 ($\sim$ 1$^{\prime}$.34 across) from \citealt{Goudfrooij1990a}, and the flux map of [NII] that we present in this study (30$^{\prime\prime}$ across). North is up and east is left in all images.\label{fig:deep images}}
    
\end{figure*}

\subsection{Extracting Kinematics}
\label{subsec:Muse Kinematics}

We used the Penalized PiXel-Fitting algorithim \citep[pPXF\footnote{Publicly available from \url{http://purl.org/cappellari/software}}, developed by][]{Cappellari2004, Cappellari2017} to calculate the line-of-sight-velocity distribution (LOSVD) of the stellar and gas components of IC 1459. To extract the stellar and gas kinematics from the MUSE cube with pPXF, we used both stellar and gas emission-line templates, which are parametrised using Gauss-Hermite functions \citep{Cappellari2017} . We used the Medium-resolution Isaac Newton Telescope Library of Empirical Spectra \citep[MILES;][]{Sanchez-Blazquez2006,Vazdekis2010}\footnote{The MILES library is available from \url{http://miles.iac.es/.}} library of stellar model templates to fit the average spectrum in each Voronoi bin.

pPXF was run on the normalised and averaged spectrum within each Voronoi bin across the cube. All spectra from MUSE and templates were logarithmically rebinned with flux conserved prior to fitting. As recommended for pPXF in \cite{Cappellari2017}, to extract the best possible fits of the stellar and gas components, we fit the spectra in two stages and the $\chi^2$ was minimised to find the best fit. We fitted the stellar and gas components simultaneously but used different fitting regions and input parameters optimised to cleanly extract information from each. 

We created a noise spectrum for each bin by normalising and median combining the squared and rooted noise spectra from the error cube. We then scaled this combined error spectrum to a more accurate measure of the S/N over the stellar fitting region by doing an initial run of pPXF to measure the "residual noise" from each binned spectrum. We fitted the spectrum and subtracted the best fit from the data to get the residuals. We then measured residual noise by taking the outlier-resistant standard deviation (1.4826  $\times$ median absolute deviation or MAD\_STD) and the signal by taking the median of each spectrum over the fitting region to determine a signal-to-residual noise (S/rN) value for each bin. The error spectra were accurately scaled to the S/rN of the fitting region determined for each bin, these values ranged from $\sim$ 70 in the central bright pixel out to the limiting binning threshold in the outer region of $\sim$ 20. The S/rN scaled error spectra were logarithmically rebinned and used as inputs to pPXF for the stellar and gas kinematic fits. 

To extract stellar kinematics, we used pPXF to fit the $\sim$ 4750--5850 $\text{\AA}$ interval for the stellar component, as it is free from strong gas emission lines and contains the strongest stellar absorption features with minimal sky emission. We masked one strong sky-line region between $\sim$ 5570--5587 $\text{\AA}$. We fit the spectrum using six moments of the LOSVD ($V$, $\sigma$, $h_3$, $h_4$, $h_5$, $h_6$) and used additive polynomials of order 10 to extract the stellar kinematics.

The gas kinematics were then fitted separately with pPXF where the previously measured stellar kinematics ($V$, $\sigma$) from the corresponding Voronoi bin were held fixed for the stellar templates only. This reduced the number of free parameters when freely fitting the gas emission templates and is as recommended in \cite{Cappellari2017}. To extract the gas kinematics, we fitted over the $\sim$ 4750--6900 {\AA} interval and excluded regions affected by strong sky-line features ($\sim$ 5570--5587 {\AA}, $\sim$ 6295--6305 $\text{\AA}$) using pPXF's \texttt{goodpixels} feature. We constructed a set of Gaussian gas emission-line templates to fit the gas emission lines in our wavelength range (H$\beta$, H$\alpha$, [SII] $\lambda$6731, 6731, [OIII]\_d $\lambda$5007, [OI]\_d $\lambda$6300, [NII]\_d $\lambda$6583). We specified the Balmer series to be input as separate templates (\texttt{tie\_balmer}=False) and the [OIII] and [SII] doublet ratios to be limited in-line with theory (\texttt{limit\_doublets}=True). The [OI], [OIII] and [NII] doublets (*\_d) use the fixed theoretical flux ratios, so only the stronger of the two lines are fitted. We fitted the gas emission lines with two moments of LOSVD ($V$, $\sigma$) and used multiplicative polynomials of order 4. 

The uncertainties of the stellar and gas LOSVD moments were calculated by performing Monte Carlo simulations for each bin. The original spectrum was perturbed 50 times, using random values selected from a standard normal Gaussian distribution (mean 0, variance 1) multiplied by each wavelength pixel of the S/rN-scaled error spectrum, to create 50 bootstrap spectra for each bin. We repeated the pPXF fits using the same input parameters for the gas and stellar fits respectively for each of the bootstrap spectra. The errors on the stellar and gas LOSVD parameters were then derived using the outlier-resistant MAD\_STD of the values from the fits of the 50 bootstrap spectra. The stellar kinematic maps and their errors were used for the orbital modelling described in Section \ref{sec:dyn}.

\begin{figure*}
\centering
\includegraphics[width=0.9\textwidth, trim=0 0 0 0, clip]{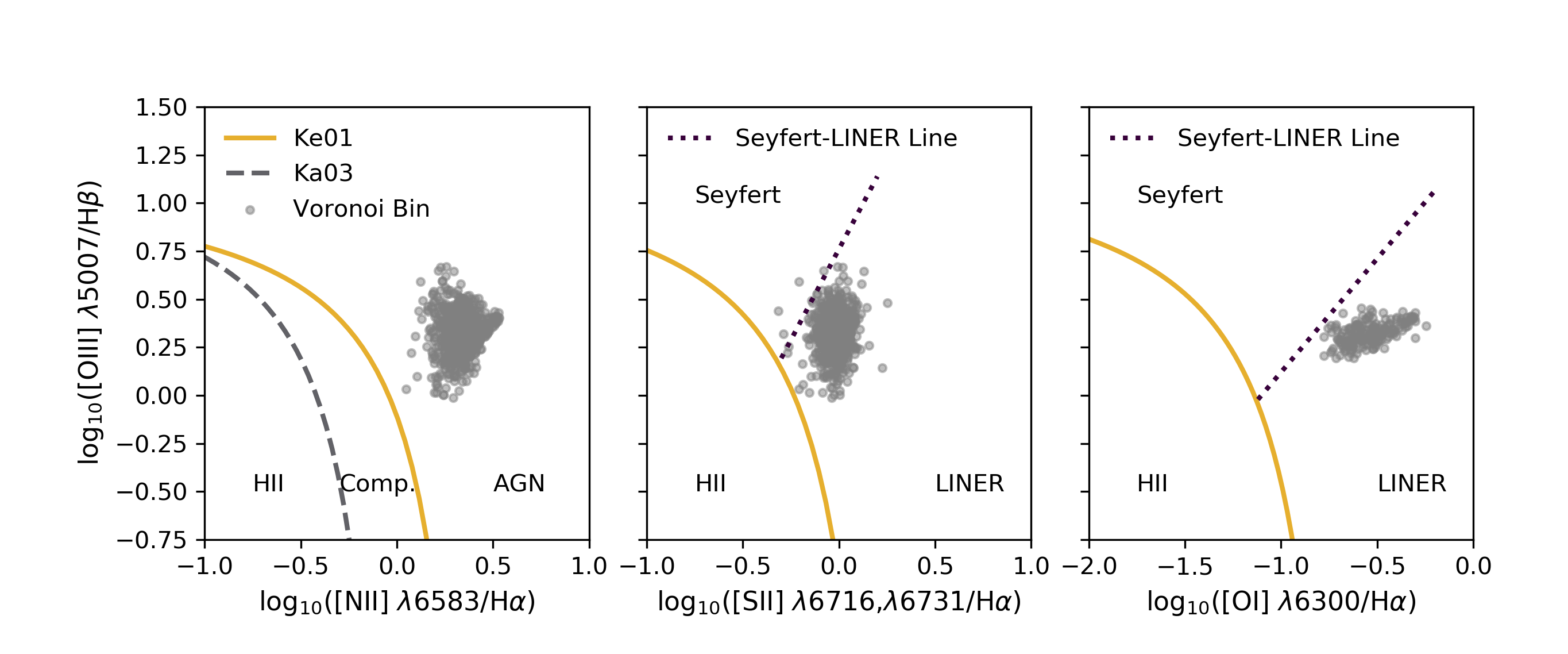}\vspace{1mm}
\caption{BPT diagrams for IC 1459. From left to right: [OIII]/H$\beta$ vs. [NII]/H$\alpha$, [OIII]/H$\beta$ vs. [SII]/H$\alpha$, and [OIII]/H$\beta$ vs. [OI]/ H$\alpha$. Each point on the diagnostic diagrams illustrates a Voronoi bin whose emission lines had an A/rN $>$ 3. The \citealt{Kewley2001} (Ke01) maximum starburst line appears on each diagnostic diagram (solid, orange line) and separates composite (below) and AGN excitation (above). The \citealt{Kauffman2003} (Ka03) pure SF line (grey dashed line), which appears on the [OIII]/H$\beta$ vs. [NII]/H$\alpha$, separates SF excitation (HII; below) and composite excitation (above). The Seyfert-LINER line from \citealt{Kewley2006} (dotted line) is in the [OIII]/H$\beta$ vs. [SII]/H$\alpha$, and [OIII]/H$\beta$ vs. [OI]/ H$\alpha$ diagnostic diagrams. The Seyfert-LINER line separates spaxels with Seyfert and LINER-like emission. See Section \ref{sec:The ionized Gas of IC 1459}. \label{fig:BPTSubPlot}}
\end{figure*}

\section{The Ionized Gas of IC 1459}
\label{sec:The ionized Gas of IC 1459}

\subsection{Maps of Ionized Gas in IC 1459}
\label{subsec:Maps of ionized Gas in IC 1459}

We present spatially resolved kinematic and flux maps of the emission-line gasses in Figure \ref{fig:GasMaps}. Only those bins with an amplitude-to-residual noise ratio (A/rN) >3 are plotted \citep[following][]{Sarzi2006}. The kinematic maps in the second column of Figure \ref{fig:GasMaps} indicate that the gas is blue shifted with a maximum relative velocity of $-$344 km s$^{-1} \pm 9$ and red shifted with a maximum relative velocity of 389 km s$^{-1} \pm 7$. In the very centre of the galaxy (2$^{\prime\prime}$), the rotation of gas is disordered and dispersion dominated. At larger radii the gas has a filamentary structure, high velocity dispersion ($\sigma_{[NII]} = $ 92 km s$^{-1}$) and gas clouds follow non-circular motions. The ionized gas in IC 1459 rotates in the opposite direction than the central stellar component (right column in Figure \ref{fig:GasMaps}) and is discussed further in Section \ref{sec:Conclusion}.    

The weakest and least abundant emission lines are H$\beta$ and [OI] and they appear to exist mostly in the very centre of the galaxy (2$^{\prime\prime}$). The detection of these weaker lines is relatively limited by the short exposure of the MUSE cube, poor seeing conditions and high illumination from the Moon during the observations. The [OIII] line is thought to arise from narrow-line regions around the AGN \citep[see ][]{Osterbrock1989}. In IC 1459, the [OIII] feature is strongest around the centre, as are most lines, which is an indication of AGN excitation in the central 2$^{\prime\prime}$. The strongest and most abundant emission line is [NII] and it exhibits a spiral structure or tidal feature, seen most prominently in the flux map (left column of Figure \ref{fig:GasMaps}), extending from the galaxy centre to the south, then east, then north region of the MUSE coverage of IC 1459.  

In Figure \ref{fig:deep images}, we show images of IC 1459 at progressively smaller radii. We show the gas-rich group environment HI map from \cite{Saponara2018}, the optical image from \cite{Malin1985} and the H$\alpha +$[NII] image of IC 1459 ($\sim$ 1$^{\prime}$.34 across) from \cite{Goudfrooij1990a}. All these have the same orientation as the MUSE maps in Figure \ref{fig:GasMaps}, and the flux map for [NII] is shown in the plot to the right for comparison. In \cite{Goudfrooij1990a} (Figure \ref{fig:deep images}, third panel from left), a clear spiral structure extends from the galaxy centre to the north east (top left) then west (right). This spiral feature sits outside of our MUSE data cube. Our results, however, focus on the central $\sim$ 1/3 of the image where structure was previously unresolved and the spiral-like structure continues through to the galaxy centre in our MUSE data. The flux in the central region of the map that we present is dominated by [NII]. The image from \cite{Goudfrooij1990a} is a composite of H$\alpha$+[NII] which means that regions towards the outskirts could be dominated by H$\alpha$ as seen in other IFU surveys \citep[e.g.,][]{Belfiore2015}.

\begin{figure}
\centering
\includegraphics[width=.49\textwidth]{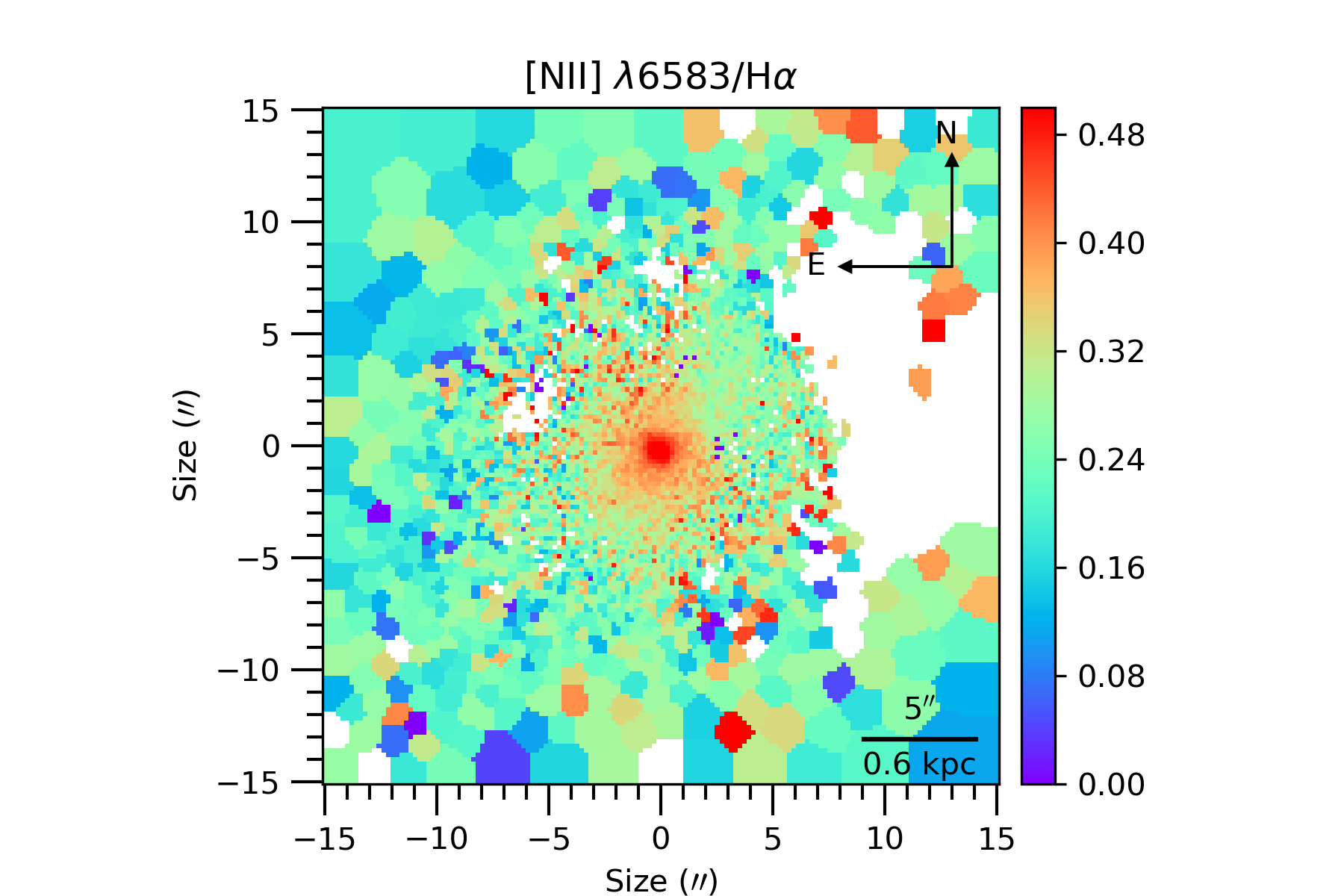}
\caption{Masked [NII] $\lambda$6583/H$\alpha$ ratio map of IC 1459. The central 5$^{\prime\prime}$ of the galaxy show higher ratios, indicative of shock excitation. See Section \ref{subsec:PowerMechanism}. \label{fig:excitationMap}}

\end{figure}

\begin{figure}
\centering
\includegraphics[width=0.42\textwidth, trim=0 0 0 0, clip]{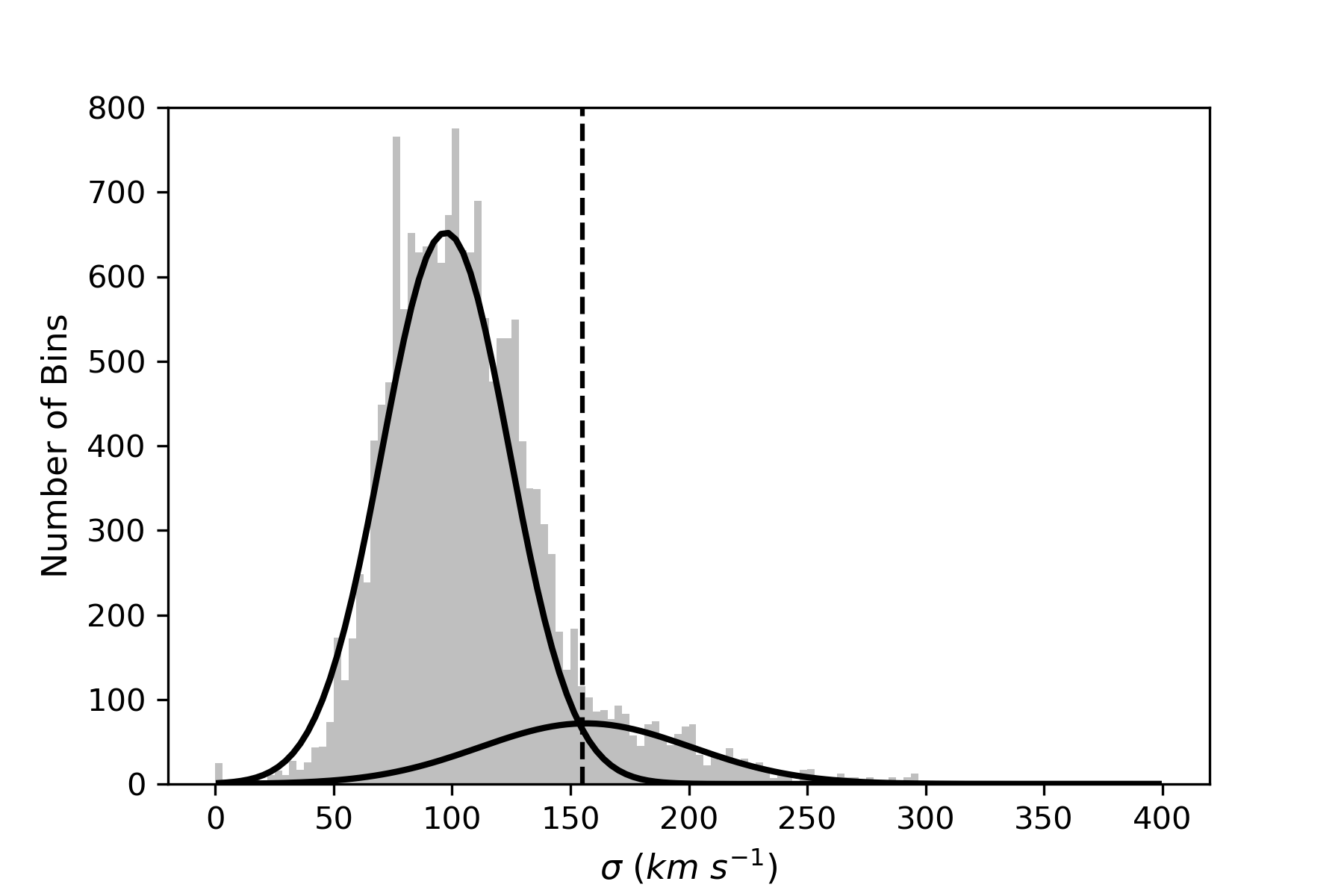}\vspace{1mm}
\caption{Histogram showing the velocity dispersion of the emission lines with an A/rN $>$ 3. We overlay best-fitting Gaussian curves (solid black lines) and show their intersection at 155 km s$^{-1}$ (dashed black line) that is used to distinguish between the broad and narrow-line components. See Section \ref{subsec:PowerMechanism}. \label{fig:VelDispHist}
}

\end{figure}

\subsection{Powering Mechanisms of Ionized Gas}
\label{subsec:PowerMechanism}

To determine the dominant power source of the emission-line gasses in IC 1459, we created the BPT diagrams \citep[shown in Figure \ref{fig:BPTSubPlot}; ][]{Baldwin1981, Veilleux1987}. Several lines of demarcation appear on these diagrams. \cite{Kewley2001} (Ke01) calculated a theoretical maximum star formation line, which is determined by pure stellar photoionization models' extrenum. In Figure \ref{fig:BPTSubPlot}, the Ke01 line appears as the solid, orange line; all data points above Ke01 are expected to be dominated by AGN activity. \cite{Kauffman2003} (Ka03) adapted the Ke01 line to include a composite region where both AGN activity and stellar photoionization are power sources for the emission-line profiles. In Figure \ref{fig:BPTSubPlot}, the Ka03 line is represented by the dashed, grey line. \cite{Kewley2006} created a new classification, which separates AGN activity from LINER and Seyfert galaxies on the [SII] $\lambda$6716, 6731/H$\alpha$ and [OI] $\lambda$6300/H$\alpha$ BPT diagrams. The \cite{Kewley2006} classification scheme is displayed as the dotted line on Figure \ref{fig:BPTSubPlot}. 

The BPT diagrams in Figure \ref{fig:BPTSubPlot} clearly indicate that the gas is not ionized by young stars. Figure \ref{fig:BPTSubPlot} reveals that the gas in IC 1459 has LINER-like emission that could either be attributed to the central nuclear region or to extended low ionization emission regions (LIER) related to IC 1459's evolved stellar population, which is observed frequently in similar galaxies \citep[e.g.,][]{Sarzi2006, Sarzi2009, Singh2013, Belfiore2016}. As the region where this ionization is present is quite large and distant from the nucleus, before making any conclusions we investigate other possible ionization processes. Because the detection of the H$\beta$ is limited, we spatially resolve the [NII]/H$\alpha$ emission-line ratio to identify potential regions of shocked gas (Figure \ref{fig:excitationMap}). In the [NII]/H$\alpha$ map, the ratios are enhanced in the central 2$-$5$^{\prime\prime}$ of IC 1459, which suggests that shock excitation may be responsible for LINER emission in IC 1459. To look into this further, we utilized the velocity dispersion of the emission-line gas to distinguish shock excitation from AGN activity. We present the histogram of velocity dispersion values in Figure \ref{fig:VelDispHist}. From Figure \ref{fig:VelDispHist}, we can infer that there are at least two physical mechanisms responsible for the ionizing the gas in IC 1459. We characterise these regions by narrow- ($\sigma$ $\leq$ 155 km s$^{-1}$) and broad-linewidths ($\sigma$ $>$ 155 km s$^{-1}$) and determined these regions by calculating the point of intersection of the two best-fitting Gaussian curves.

\begin{figure*}
\centering
\includegraphics[width=.95\textwidth, trim=0 0 0 0, clip]{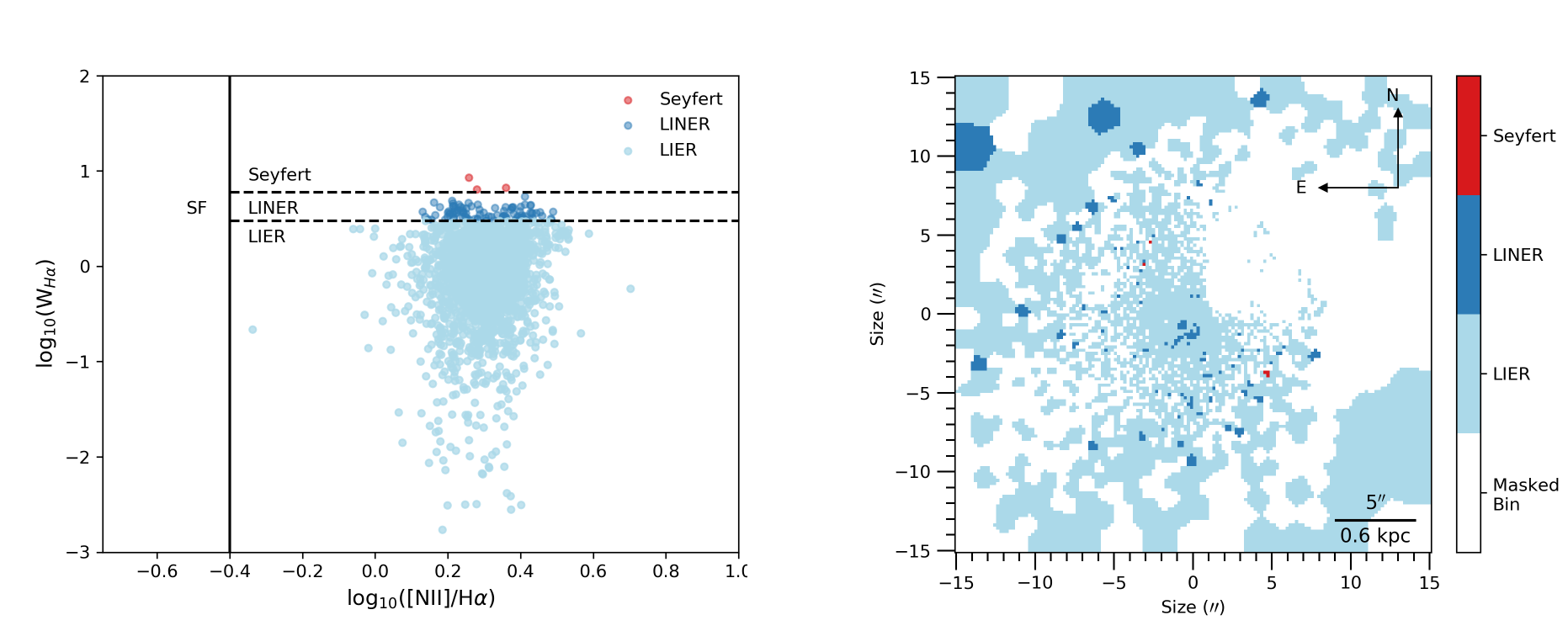}\vspace{1mm}
\caption{\textit{Left:} The WHAN diagram of IC 1459. Each point on the diagnostic diagram represents a Voronoi bin.The top most dashed line at W(H$\alpha$) $=$ 6 $\text{\AA}$ separates Seyfert (above) and LINER (below) excitation and the bottom dashed line at W(H$\alpha$) $=$ 3$\text{\AA}$ separates the LINER-type emission (above) and the LIER-type emission (below). The vertical line at [NII]/H$\alpha$ $=$ $-$0.4 separates starburst (left) and AGN (right) activity. \textit{Right:} Map showing the spatial location of the ionizing mechanisms determined from the WHAN diagram (left). See Section \ref{subsec:PowerMechanism}. \label{fig:WHAN}}

\end{figure*}

To further understand the ionizing mechanisms in IC 1459, we use the equivalent width of H$\alpha$ (W(H$\alpha$)) versus [NII]/H$\alpha$ (WHAN diagram) developed by \cite{CifFernandes2011}. In Figure \ref{fig:WHAN}, $\sim$ 2.5$\%$ of the Voronoi bins have spectra with LINER-like emission. The spaxels with W(H$\alpha) <$ 3$\text{\AA}$ have emission related neither to AGN nor to starburst activity. Rather, the ionizing mechanism for the gas in these spaxels is most likely from post-asymptotic giant branch (AGB) stars, and can be characterised as LIER-like. From Figure \ref{fig:WHAN}, LIER excitation is the dominant powering mechanism in IC 1459 and LINER excitation is mostly limited to the central 2$^{\prime\prime}$. This centrally concentrated LINER excitation shows hints of filamentary or jet-like structure. Three bins exhibit Seyfert-like emission, but from their non-centralized and random locations it is evident that these are likely spurious detections and should be ignored.

We speculate that the ionized gas of the narrow component can be explained by the accretion of gas from the intra-group medium, and is therefore the result of tidally disrupted gas from group interactions \citep{Serra2015, Saponara2018}. This is supported by the extended spiral-like feature we find within the centre of the galaxy and those tidal features found at significantly larger radii in other studies out to gas reservoirs in the larger group environment \citep[Figure \ref{fig:deep images}; ][]{Malin1985, Goudfrooij1990a, Iodice2020}. The ionized gas of the broad line-width component ($\sigma$ $>$ 155 km s$^{-1}$), however, has two coexisting heating mechanisms according to Figure \ref{fig:WHAN}: LINER emission from accretion onto the AGN and LIER emission from post-AGB stars. 
 
One remarkable feature of the accreted ionized gas is that it is counter-rotating with respect to the KDC, which implies that ionized gas is not related to the KDC. A timing argument can be used to support this conclusion. If we assume that the HI gas in the IC 1459 group is tidally stripped from one of the gas-rich galaxies, then this is likely a relatively recent event. \citet{Oosterloo2018} proposes 2 Gyr as a lower limit. This can be compared with the estimated survival time of $\approx 4$ Gyr of the tidal features around IC 1459 \citep{Mancillas2019,Iodice2020}, but is significantly younger than the age of the stellar populations found in the KDC, and IC 1459 as a whole \citep[7.6--9.6 Gyr;][]{Prichard2019}. Further evidence that the gas is not related to the formation of the KDC can be established by looking at the stellar orbital distribution of IC 1459.

\section{Stellar orbits within IC 1459}
\label{sec:dyn}

In addition to the prominent KDC in the stellar velocity map, MUSE observations of IC 1459 revealed a somewhat unusual feature in the velocity dispersion map \citep[Figure~\ref{fig:GasMaps},][]{Prichard2019}. Next to the central peak, there is an elongated plateau of high velocity dispersion values, which stretches along the major axis and beyond the visible part of the KDC to about 10\arcsec\,from the centre. The velocity dispersion is relatively smaller (still above 200 km/s) along the minor axis of the galaxy, while high ($\sim300$ km/s) along the major axis and subsequently ``flares" away from the major axis at radii larger than 10\arcsec\,(at the edge of the MUSE field). The high velocity dispersion along the major axis is also noticeable on \citet{Franx1988} long-slit data, but the MUSE maps of \citet{Prichard2019} reveal for the first time the true morphology of the feature (see Figure~1 of \citet{Cappellari2002} for an interpolated velocity dispersion map based on long-slits observations). 

Such a velocity feature is reminiscent of the $2\sigma$ features visible in the velocity dispersion maps of some disk galaxies \citep{Krajnovic2011}. They are characterised with two symmetric peaks along the major axis, which arise as these galaxies are made of two counter-rotating disks \citep[e.g. NGC4550, as a prototypical example of the class;][]{Rix1992, Rubin1992, Emsellem2004}. This suggests that IC 1459 could also have strong counter-rotating components, related to the KDC. IC 1459 is not a disk galaxy, nor it is likely made of two counter-rotating disks. Nevertheless, it is analogous to another giant elliptical, NGC 5813, which has a similar high velocity dispersion plateau \citep{Krajnovic2015}. Dynamical models showed that stellar kinematics of NGC 5813 can be reproduced by two counter-rotating orbital families, one of which is dominant in the centre and is responsible for the KDC, while at larger radii the two families contribute with similar masses, cancelling out the net rotation, but increasing the line width (dispersion) along the major axis \citep{Krajnovic2015}. The similarities between the kinematics of IC 1459 and NGC 5813 are striking, and we set up dynamical models to investigate the internal structure of IC 1459.

\begin{figure*}
    \includegraphics[width=\textwidth]{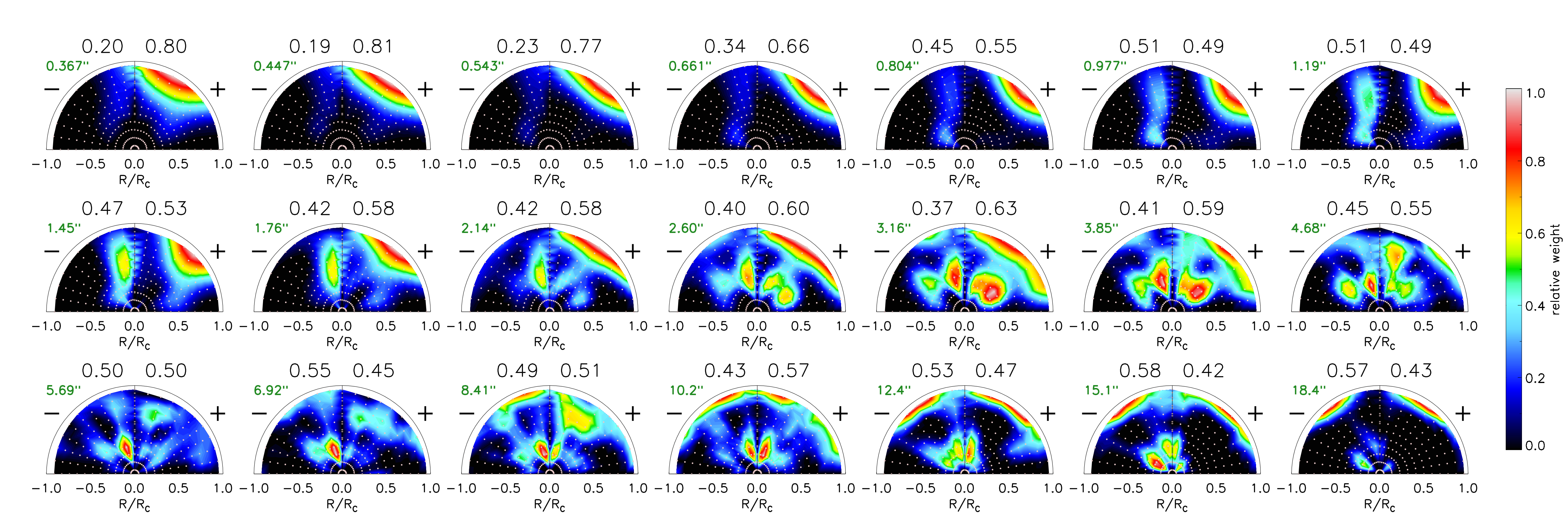}
    \caption{Orbital space of the axisymmetric Schwarzschild model of IC 1459 represented with weights (in colour) assigned to orbits that bulid the model. Each orbit is defined by three integrals of motion: energy ($E$), projections of the angular momentum along the short axis ($L_z$) and the non-analytic third integral. Each hemisphere plots the meridional plane ($R$,$z$) for a given energy, where the white dots indicate the initial positions of orbits. Green numbers in the upper-left corner show the radius (in arcsec) of the circular orbit of a given energy. Within each hemisphere, the angular momentum changes along the $x$-axis while the value of the third integral varies along the $y$-axis. Prograde orbits are shown on the right side of each panel (indicated with a ``$+$'' sign), while the retrograde orbits are shown on the left side (``$-$'' sign). The ratios of the total mass assigned to prograde and retrograde orbits are printed above the corresponding side of the panel. We show only a subset of radii, representative of the observed stellar kinematics. The relative colour coding of the orbital mass weights is shown with the colourbar, where 1 represents the largest mass weight assigned to an orbit at the given energy. \label{fig:StOrbitModel}}
    
\end{figure*}

We constructed axisymmetric Schwarzschild models based on the orbit superposition approach \citep{Schwarzschild1979} using the Leiden code \citep{Cappellari2006}. Based on an assumption on the gravitational potential, the code builds a library of orbits, which are then combined to reproduce the observed stellar distribution and kinematics. Axisymmetric gravitational potentials support only one family of stable orbits, the short axis tubes \citep{deZeeuw1985}, which are defined by three integrals of motion: energy ($E$), the short axis projection of the angular momentum ($L_z$), and the third (non-analytical) integral ($I_3$). This galaxy was modelled already with an earlier version of the code \citep{Cappellari2002}, but there are some important differences: here we use an evolved version of the code adapted for integral-field data (in terms of the orbital set up) and a much higher number of orbits, as specified above. The biggest difference, however, is in the MUSE stellar kinematics used to constrain the models. The details of Schwarzschild modelling and our set up are given in Appendix~\ref{a:schw}. 

In Figure ~\ref{fig:StOrbitModel} we show the main result of the modelling: meridional planes of the galaxy, represented by polar grids of orbital starting position (white dots), each panel is for a given energy (represented with the radius of the circular orbit at that energy shown in green). Each panel is divided into right and left sides, corresponding to prograde and retrograde orbits, respectively. The definition of what is prograde and retrograde is arbitrary, and we assign the component responsible for the KDC as the prograde component (``$+$" sign). High angular momentum orbits are found in the lower right and left corners of the panels. Orbits with increasing $I_3$ values are found higher along the $y$-axis. The colour contours show the weights assigned by the model to the orbits. Not all orbits are selected (black regions), and there are regions in the meridional plane which are given more weight. 

The conclusion is that the model requires both prograde and retrograde orbits throughout the galaxy body, each component having essentially the same fraction of the total mass (51:49 per cent split, within the extent of the MUSE data). Within the KDC (for $r<5\arcsec$) the ratio of the prograde to retrograde orbits is approximately 58:42 per cent, this is followed by a transition region between 6\arcsec and 12\arcsec with the orbital split of 50:50, while at the large radii ($r>12\arcsec$) this ratio reverses to 43:57. Therefore, the KDC in IC 1459 is not a separate entity residing in the centre, kinematically decoupled from the rest of the galaxy. On the contrary, the KDC is an integral part of the galaxy made up of orbits belonging to a single family of short axis tubes that rotate in the prograde sense, and extend from the central parts to large scales. There is another family of short axis orbits which also extend throughout the system, but rotate in the retrograde sense. The morphology of the velocity map (the KDC and its counter-rotation with respect to the large scales) can be explained through a superposition of these two orbital families, where the prograde dominates in the central parts ($<5-10\arcsec$), while the retrograde in the outer parts ($>5-10\arcsec$). 

\citet{Cappellari2002} also presented the orbital structure of their best fit model, which bears many similarities to our model. They do not provide details on the mass fractions of the prograde and retrograde orbital families, but the model implies a more localised presence of the prograde orbits, which are mostly restricted to the KDC region. The difference in the results between their and our work originates essentially from the improved quality of our data. 

The continuous distribution of both prograde and retrograde orbits within the area covered by the MUSE data also supports conclusions of stellar population analysis from \cite{Prichard2019}: IC 1459 has relatively constant initial mass function (IMF) and metallicity, the galaxy comprises a fairly homogeneous stellar population, and the apparent counter-rotating core is due to a slight mass imbalance of stars on orbits. 

From a dynamical point of view, the two orbital families have to come from different progenitors of similar masses. This major merger formed the central regions (observed by MUSE) of the galaxy and is responsible for the observed stellar kinematics. The emission-line gas might have a similar angular momentum vector as the retrograde component that is dominating the outer parts of the galaxy, but this is not sufficient to associate it to the formation of this or any other stellar component. Based on the timing argument for the accretion of gas, the old stellar populations, as well as taking into account the equal masses of the prograde and retrograde components, the emission-line gas is not related to the formation of stars making up the current stellar distribution. Therefore, the ionized gas in IC 1459 must come from a later-stage accretion. 

\section{Discussion and Conclusion}
\label{sec:Conclusion}

In this work we have investigated the origin of ionized gas in IC 1459, a massive, slow-rotator ETG with a rapidly counter-rotating stellar core. We used IFU data from MUSE to study spatially resolved structure, flux, and kinematics of the gas emission lines. We also investigated the orbital space of the stellar populations through Schwarzschild modelling. This combined study of both the ionized gas and stars in IC 1459 builds on previous work to construct a more complete evolutionary history of the galaxy. 

We used the Voronoi binning method of \cite{Cappellari2003} to adaptively bin each spaxel over the MUSE cube to a roughly constant signal to noise (S/N $>$ 20). Using pPXF, we extracted the stellar and gas information from the MUSE data cube and assessed their spatially resolved structure. We determined the dominant power source of emission-line gasses in IC 1459 by evaluating emission-line gas maps, diagnostics, and velocity dispersions. We also inspected the stellar orbital distribution through Schwarzchild modelling. Our key findings are outlined below. 

\begin{enumerate}
    \item We present the first detailed flux and kinematic maps of  the ionized gas in the central 30$^{\prime\prime}$ $\times$ 30$^{\prime\prime}$ of IC 1459. The gas is blue shifted to the south east with a maximum relative velocity of $-$344 km s$^{-1} \pm 9$ and red shifted to the northwest with a maximum relative velocity of 389 km s$^{-1} \pm 7$. The gas in the central 2$^{\prime\prime}$ of the galaxy has disordered, dispersion-dominated kinematics.
    
    \item The least abundant emission lines are H$\beta$ and [OI] and we find these exist mostly in the central 2$^{\prime\prime}$ of the galaxy. The strongest, most abundant emission line is [NII], which exhibits an extended disk-like structure or tidal feature within the MUSE coverage of IC 1459. The structure of the [OIII], H$\alpha$ and [SII] emission-line gasses appears to be in a spiral-like structure along the galaxy's major axis.
    
    \item We find that LINER and LIER excitation are the two dominant ionizing mechanisms in IC 1459. By investigating the velocity dispersion, emission-line ratios and the WHAN diagram of the emission-line gas, we determined that the ionization mechanisms could have contributions from shock heating and AGN excitation. However, given that LIER-like emission in IC 1459 is extended (kpc scale), the gas is most likely heated by evolved AGB stars within the old stellar core rather than a central nuclear source. This result adds to a growing body of literature that extended LIER emission is from pAGB stars rather than a central, low luminosity AGN  \citep[e.g.; ][]{Sarzi2006, Singh2013, Belfiore2016}.
    
    \item The central structure of ionized gas is rotating in the opposite direction to the central stellar component, suggesting that the origin of the ionized gas is different than that of the central stellar component. Although it is likely that IC 1459 has undergone at least one gas-rich merger in its past, we do not believe the ionized gas currently present in IC 1459 is from a major merger event with a spiral galaxy from the group of galaxies surrounding IC1459. Within the IC 1459 galaxy group, there is evidence for a large \citep[500 kpc-long; ][]{Oosterloo2018}, massive \citep[10$^{10}$ M$_{\odot}$; ][]{Iodice2020} HI debris complex near IC 1459  \citep{Saponara2018}. We propose that the ionized gas that we detect in the central 30$^{\prime\prime}$ $\times$ 30$^{\prime\prime}$ of IC 1459 is from the ongoing, late-stage accretion from the group environment. Given the extent of the HI debris complex, the velocity dispersion of the IC 1459 group \citep[223 $\pm$ 62 km s$^{-1}$; ][]{Brough2006}, along with simulations from \cite{Mancillas2019}, the accretion onto IC 1459 has been occurring for the last $\approx$ 2$-$4 Gyrs \citep[Figure \ref{fig:deep images};][]{Oosterloo2018, Iodice2020}.
    
    \item The stellar content of IC 1459 is evenly split into a prograde and a retrograde component. The prograde component is somewhat more centrally concentrated, while the retrograde component dominates at larger scales.  Their (luminosity weighted) superposition is responsible for the complex appearance of the velocity and velocity dispersion maps. This also means that the origin of the KDC in IC 1459 is linked to an early major merger, and is not related to more recent interaction events, evident in the tidal features and the abundance of the inter-group gas. 

\end{enumerate}

\section*{Acknowledgements}
We thank the anonymous referee for valuable comments on the paper. CRM and RAJ gratefully acknowledge support by NSF REU grant AST-1757321, the Massachusetts Space Grant, and by the Nantucket Maria Mitchell Association.

\section{Data Availability}
The reduced MUSE data that support the findings of this study are available in the ESO Science Archive at \url{http://archive.eso.org/cms.html} and can be accessed with program ID 094.B-0298(A).

\bibliographystyle{mnras}
\bibliography{IC1459bib.bib}

\appendix

\section{A Schwarzschild model of IC1459}
\label{a:schw}

We use the Leiden version of the Schwarzschild orbital superposition code as presented by \citet{Cappellari2006}. The Schwarzschild codes are typically used to determine the masses of supermassive black holes in the centres of galaxies, but in this case we are only interested in the global distribution of orbits in IC1459. As mentioned in the main text, there is only one orbital family, short axis tubes, which can rotate in the prograde and retrograde sense within an axisymmetric potential well. Each orbit is defined by three integrals of motion: energy ($E$), the short axis projection on the angular momentum ($L_z$), and the third integral ($I_3$). The code samples the orbits on 41 energy points, which are radially spaced throughout the galaxy with a logarithmic step. In addition, for each radius defined by the energy step, there are a further 11 angular and 11 radial points, arranged in a polar grid, which sample the combination of $L_z$ and $I_3$. The orbits are also dithered (6 dithers in each dimension), so the final numbers of orbits superseeds 2.1 million \citep[for details see][]{Cappellari2006}. In order to ensure a smooth distribution of orbital weights we apply a moderate regularisation \citep[$\Delta=4$, see ][]{vanderMarel1998}, but we verified that the results did not differ significantly when the regularisation was not used.

As we are interested only in the internal orbital structure we use some of the input parameters from \citet{Cappellari2002}, which modeled IC1459 to estimate the mass of the supermassive black hole. We use their parameterisation of stellar light (their Table 4), inclination ($90\deg$), mass of the central black hole (M$_{BH}=2.6\times10^{9}$ M$_\odot$), and the stellar mass to light ratio (M/L=3). In this way, the gravitational potential is fully specified and only the MUSE kinematics are used to constrain which orbits will end up in the final model representing the galaxy. 

IC1459 has a small photometric twist and, being a giant elliptical and a massive slow rotator, it is likely that it is at least weakly triaxial \citep{Weijmans2014}. However, the central kinematics covered by MUSE are largely bisymetric with respect to the (projected) minor axis of the galaxy, although a kinematic twist is visible towards the edge of the field. Nevertheless, as we focus on the central regions, we follow \citet{Cappellari2002}, and apply the axisymmetric code as an approximate dynamical models \citep[similar to assumptions for modelling NGC 5813 by][]{Krajnovic2015}. As the models are by construction axisymmetric, we symmetrise the MUSE kinematics using point-(anti)symmetry around the photometric minor axis (PA=$123\degr$), averaging the values at positions: $[(x,y),(x,-y),(-x,y),(-x,-y)]$, but keeping the original errors for each point.

\begin{figure*}
    \includegraphics[width=\textwidth]{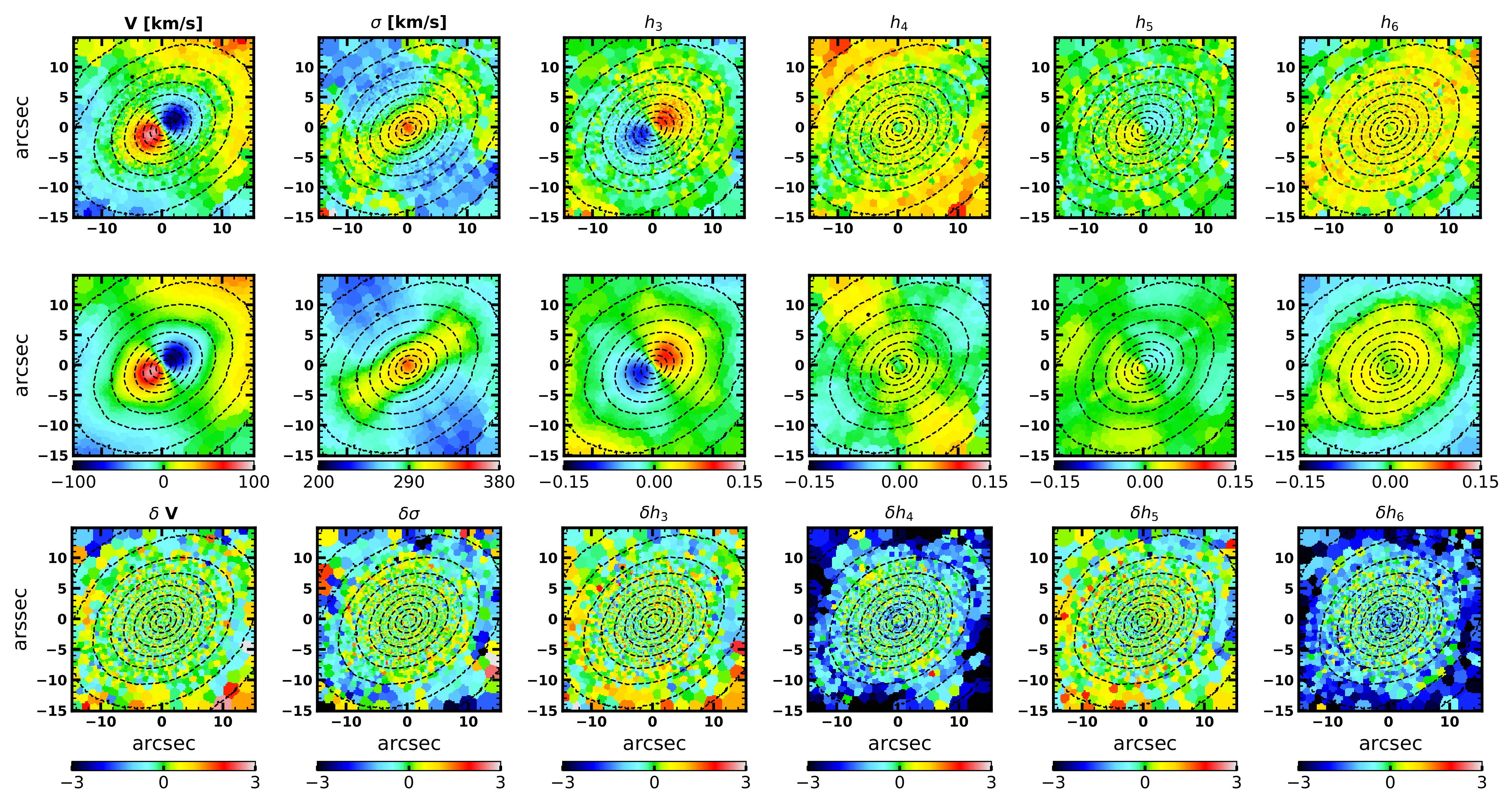}
    \caption{Comparison between the kinematic data and the result of the Schwarzschild modelling. {\it Top row:} Symmetrised MUSE stellar kinematics, starting from left to right: the mean velocity, the velocity dispersion, and the Gauss-Hermite coefficients $h_3$, $h_4$, $h_5$ and $h_6$. {\it Middle row:} kinematics of the Schwarzschild dynamical model, showing the same maps as in the top row. {\it Bottom row:} Residuals between the Schwarzschild models and the symmetrised MUSE kinematics, divided by the observational error. \label{fig:SchwModel}}
    
\end{figure*}

In Figure~\ref{fig:SchwModel} we show the quality of our Schwarzschild model in the sense of the kinematic data-model prediction comparison. The bottom row in the figure shows the model$-$data residuals, divided by the errors. The model is able to reproduce the features on the velocity and the velocity dispersion maps  well, as well as features on odd Gauss-Hermite coefficient maps. The biggest and systematic discrepancies are found for $h_4$ and $h_6$ maps where model over-predicts the data; they are however within $2 \sigma$ uncertainty level except in a few bins towards the edges of the field.

\bsp	
\label{lastpage}
\end{document}